# VARIATIONAL PRINCIPLES IN CHEMICAL EQUILIBRIA: COMPLEX CHEMICAL SYSTEMS WITH INTERACTING SUBSYSTEMS

B. Zilbergleyt[I]

INTRODUCTION

The goal of this paper is to formulate a revised condition of global equilibrium in complex chemical systems as variational principle, using formalism of recently developed discrete thermodynamics of chemical equilibria (DTD) [1,2,3].

Equilibrium in chemical systems is usually understood as "true" thermodynamic equilibrium (TdE) with stationary variables and minimal free energy, by virtue of its premises true only for isolated systems. Notion of equilibrium, however, demands any balance to be achieved between more than one counteracting entities or components for no one entity can interact with and balance itself. It is uncertain, what one means by equilibrium until the balance defining factors are specified. From this point of view, classically defined "internal" thermodynamic equilibrium as a mono-balance is too abstract and makes sense neither semantically nor by essence. Unlike classical thermodynamics, DTD defines equilibrium as a balance between external and internal thermodynamic forces (TdF), acting against chemical system; such an approach is more general and covers various kinds of stationary states [4].

Thermodynamic forces, external to a non-isolated system, may be of two kinds: forces of a non-chemical origin, acting without material exchange, like in electrochemical or photochemical cells (closed system), and the forces, imposed upon the open subsystem by other chemical subsystems of the mother system and acting with material exchange. Non-isolated systems must be considered as subsystems of a larger system[II]; here we observe equilibrium between the subsystem and its environment, or a balance between external and internal TdF. If the system is isolated, in order to define its thermodynamic equilibrium it should allow subdivision by smaller entities (in the set-builder notation S={s: *s is a subsystem*}), mutually balanced by thermodynamic forces. Now general condition of equilibrium for chemical systems is

(1) $$F_{je} + A_{ji} = 0,$$

where $F_{je}$ is a resultant of external thermodynamic forces, acting against *j*-system, and $A_{ji}$ is the bound affinity, an internal thermodynamic force, resisting to the external forces and mirroring them [1]. In isolated system $F_{je}=0$, and equation (1) turns to $A_{ji}=0$, matching the classical result.

DTD defines thermodynamic functions and parameters in finite differences; e.g., thermodynamic affinity is $A_j = -\Delta G_j/\Delta\xi_j$ with reaction extent $\Delta\xi_j$, a chemical distance between the initial point and the current state of the system; in TdE $\Delta\xi_j$ equals to unity. Therefore the shift from TdE, a result of the external thermodynamic impact and a measure of the subsystem deviation from TdE, is $\delta\xi_j = 1 - \Delta\xi_j$. Further on in writing we use $\Delta_j$ for $\Delta\xi_j$ and $\delta_j$ for $\delta\xi_j$. The shift is the only variable of the theory, its introduction creates a new reference frame for the states of chemical system, in which TdE rests at $\delta_j=0$ (a tautological definition of TdE), or $\Delta_j=1$. At p,T=const, equation (1) leads to equilibrium condition for a non-isolated chemical subsystem

(2) $$\Sigma_j[F_j - \Delta G_j(\delta_j)/\Delta_j] = 0.$$

Taking into account the Le Chatelier Response [6], which sets a functional relationship between external TdF and caused by it system shift from TdE, the unfolded form of (2) may be written down as a logistic map of states of the j-system

(3) $$\ln[\Pi_j(\eta_j,0)/\Pi_j(\eta_j,\delta_j)] - \tau_j\varphi(\delta_j,\pi) = 0,$$

Here $\Pi_j(\eta_j,0)$ (i.e. $\delta_j=0$) and $\Pi_j(\eta_j,\delta_j)$ are the mole fraction products for the reaction, running within the system; $\eta_j=\Delta^*n_{kj}/\nu_{kj}$ is the *equivalent of chemical transformation*, a ratio between amount of k-participant $\Delta^*n_{kj}$, that must be transformed in the chemical reaction within j-subsystem in its isolated state *ab initio* to TdE, given initial composition of the system, and its stoichiometric coefficient; $\tau_j$ is the growth factor for $\delta_j$; $\pi$ is loosely defined factor of system complexity; and the force function $\varphi(\delta_j,\pi)$ equals either to $\delta_j(1-\delta_j^\pi)$ for a strong system or $(1-\delta_j^{\pi+1})$ for the weak system (not a reaction!). Detailed explanation of the terms and derivation of (3) are given in [1,2]. Consider an isolated system; in classical paradigm its equilibrium composition can be found, minimizing lagrangian [7]

(4) $$L(n, \lambda) = G(n) - \sum_{1\to N_e}\lambda_l(\alpha_i - \sum_{1\to N_s}M_{ij}n_j),$$

where $N_e$ is the number of chemical elements and $N_s$ is the number of chemical species in the system, $\alpha_i$ are the entries of the atomic element abundance vector $\alpha$, $M_{ij}$ are the entries of the molecular formula vector $M$. The second term is a constraint, imposed on the system Gibbs' free energy due to restricted amounts of chemical elements, $\lambda_l$ is a logistic Lagrange coefficient. Solution to (4) yields the minimum of $G(n)$ and the system equilibrium composition in moles *n* of chemical species. Such an approach considers a complex system as a mere sum of its species, which actually implicitly represent the hidden subsystems; the second term in (4) works just like an accountant, guarding the chemical system against overconsuming its chemical elements.

---

[I] System Dynamics Research Foundation, Chicago, USA, sdrf@ameritech.net
[II] Probably, Hertz was the first to word this idea clearly for mechanical systems as early as in year 1894, in the original publication of [5].

DTD considers the above system to be explicitly cemented by its internal interactions, and complex equilibrium in such a system is a set of closely associated open equilibria between each of the subsystems and the others, comprising its environment, or compliment to the mother system. The logistic constraints are exactly the same as in (4), but, as opposite to classical approach, we also have to take into account the constraints, caused by interactions between the subsystems. We will find them out, analyzing d'Alembert's principle and principle of virtual work in application to thermodynamics of chemical equilibria.

D'ALEMBERT'S PRINCIPLE

In mechanics traditional form of d'Alembert's principle for a system of material points is [8]

(5) $$\Sigma_j(F_j - m_j\alpha_j)\delta r_j = 0,$$

$F_j$ are the active (external) forces, $m_j$ and $\alpha_j$ are the particle masses and accelerations, their products are the forces of inertia, created by motion; $\delta r_j$ are the virtual displacements. In chemical systems one may substitute virtual displacements by virtual shifts; multiplying (2) by $\delta_j$, we get

(6) $$\Sigma_j[F_j - \Delta G_j(\delta_j)/\Delta_j]\delta_j = 0.$$

Expression (6) not only by its form, but also by its essence, with *virtual deviations* from TdE as virtual displacements matches d'Alembert's principle. Indeed, the first term in map (6) represents the resultant of active thermodynamic forces, while the second, the *bound affinity*, resisting the changes to the system state, plays the role of *thermodynamic force of inertia*[1] by virtue of its origin. To make it clear, suppose a chemical system, resting in TdE with $F_j=0$, $\delta_j=0$, $\Delta G_j(\delta_j)=0$ and also $A_j(\delta_j)=0$. At a certain moment of time the external forces become active, and the system deviates from TdE, generating the bound affinity. The latter mirrors the resultant of external forces $F_j$ and equals to it in order to keep the subsystem in equilibrium with its environment. Being originated from the system movement out of TdE, this internal force will drive the system back to TdE when $F_j$ vanishes.

Using (3) we arrive at unfold form of (6)

(7) $$\Sigma_j\{\ln[\Pi_j(\eta_j,0)/\Pi_j(\eta_j,\delta_j)] - \tau_j\varphi(\delta_j,\pi)\}\delta_j = 0,$$

The terms within braces are the above mentioned logistic maps of the subsystem states, their graphical solutions are pitchfork bifurcation diagrams in coordinates $\delta_j$ vs. $\tau_j$.

---

[1] Although some authors have used "inertia term" in thermodynamic forces (e.g. [9]), any solid and commonly recognized definition of the thermodynamic force of inertia seemingly doesn't exist.

Like equation (5), that creates essential difference between newtonian, vectorial mechanics and analytical mechanics by accounting for the forces of inertia and thus redefining the equilibrium conditions, map (6) creates a difference between the conventional paradigm, which knows only affinity as thermodynamic force, vanishing at equilibrium, and discrete thermodynamics of chemical equilibria, where chemical equilibrium exists as a balance of non-vanishing thermodynamic forces. Thermodynamic version of d'Alembert's principle may be worded as:

*any state of open or closed thermodynamic system may be considered equilibrium, if thermodynamic forces of inertia are added to the external thermodynamic forces.*

In mechanics, d'Alembert's principle reduces dynamic tasks to static and eventually gives a complete solution to problems of mechanics [5]; *in thermodynamics it offers equilibrium solutions to non-equilibrium tasks*. One new important opportunity, rendered by d'Alembert's principle to thermodynamics, is extended capability to cover not only TdE, but also equilibrium and non-equilibrium steady states. Imagine chemical reactor with in and out flows; at appropriate flow rates chemical system within such a reactor stays in a non-equilibrium steady state. Now full set of thermodynamic forces differs from the equilibrium stationary case by additional inertial forces, directly related to the fluxes; their sum can be traditionally expressed as [9]

$$(6) \qquad X_j = \Sigma_k R_{kj} J_k,$$

where $X_j$ is the resultant of the inertial forces, related to fluxes $J_k$ through the *j*-subsystem borders with the resistances $R_{kj}$. Now one can put down full expression of d'Alembert's principle for stationary states of chemical systems

$$(8) \qquad \Sigma_j (F_j - \Delta G_j(\delta_j)/\Delta_j - \Sigma_k R_{kj} J_k) \delta_j = 0.$$

The flows may be also related to mass transfer between open subsystems.

As it was mentioned by Lanczos [10], other variational principles in mechanics "… are merely mathematically different formulations of d'Alembert's principle". The same situation should be expected in thermodynamics.

PRINCIPLE OF VIRTUAL WORK

Principle of virtual work, or principle of virtual displacements [8]

$$(9) \qquad \Sigma_j F_j \delta r_j = 0$$

clearly follows from d'Alembert's principle and is an alternative definition of equilibrium state in a mechanical system with no motion and therefore in absence of the inertia forces: *mechanical system will be in equilibrium if, and only if the total virtual work of all active forces vanishes*. By d'Alembert's own words, it is "*the law of the live forces conservation*" (quoted by [11]). One can easily deduct the thermodynamic version of this principle from (3) as

$$(10) \qquad \Sigma_j F_j \delta_j = 0$$

with subsystem shifts from TdE as virtual displacements. Appropriate wording of the principle is:

*any open thermodynamic system will be in thermodynamic equilibrium if, and only if the total virtual work of all the external thermodynamic forces vanishes.*

Unlike a system of material points, where the inertia forces vanish at zero accelerations, in thermodynamic system the inertia force, or bound affinity takes on a non-zero value as soon as the system deviates from TdE and we observe non-zero shifts. With this notion, principle of virtual work (10) in combination with d'Alembert's principle (6) gives us a condition of the system equilibrium via subsystem interactions as

$$(11) \qquad \Sigma_{1 \to s} A_j \delta_j = 0.$$

This is an alternative and relatively simple form of the principle of virtual work in thermodynamics of chemical equilibria. Notice that dimension of products under the sum sign in (11) is energy due to dimensionless $\delta_j$.

CONDITION OF COMPLEX CHEMICAL EQUILIBRIUM AS VARIATIONAL PRINCIPLE

Now we have two constraints, turning the task of unconditional minimization of the system Gibbs' free energy into a task for relative minimization [12]. Let's recall that the number of *i*-atoms in j-molecules (born in j-chemical subsystem) is

(12) $$n_{ij} = \nu_{ij}\eta_j(1-\delta_j),$$

and that $A_j = -\Delta G_j/(1-\delta_j)$ and $A_j\delta_j = -\Delta G_j\delta_j/(1-\delta_j)$. Now we can put down the sought lagrangian in the matrix form with both conservation lagrangian products

(13) $$L(\delta,\lambda_l,\lambda_i) = G(\delta) - \lambda_l[\mathbf{a}-\mathbf{\nu}\eta(1-\delta)] - \lambda_i\Sigma_{1\rightarrow s}[\Delta G\delta/(1-\delta)],$$

where $\lambda_l$ is the vector of logistic Lagrange coefficients, $\lambda_i$ is the interactive Lagrange coefficient, $\delta$ is the shifts vector. Vectors $\mathbf{a}$ and $\mathbf{\eta}$ are the task parameters, defined by the system initial composition and $\Delta G^0$ of the subsystem chemical reactions (or by $\Delta G^0$ of real or potential species in the system). Minimization of lagrangian (13) gives the sought solution to the complex equilibrium task in terms of the subsystem shifts from their individual TdE. Parameter $\eta_j$, easily obtainable from simple TdE simulation for individual subsystems in their isolated states [1], allows us to recalculate equilibrium composition of complex system to moles of chemical species by the formula

(14) $$n_k = n^0_k + \Sigma_j \nu_{kj}\eta_j(1-\delta_j).$$

The first lagrangian product in (13) may be replaced by equivalent expression with the number of moles without $\delta_j$, but we cannot get rid of $\delta_j$ in its third term – it is essentially a product of discrete thermodynamics.

Recall that graphical solutions to map (2) are pitchfork bifurcation diagrams [1,2], where the subsystem state is represented by a point before bifurcation limit and a combination of two points in the bi-stability area beyond bifurcation limit. So, there may be single-root or two-root solutions (or even more) for some individual subsystems, depending upon how strong they are impacted by the counterparts, and, therefore, upon the subsystem location on its own bifurcation diagram in global equilibrium. In some cases the subsystems may show oscillatory behavior [13], able to trigger complicated oscillations of the system chemical composition in an avalanche style.

CONCLUSION

Besides finding some practical relationships to be used in thermodynamic analysis and simulation of chemical systems, in this paper we tried to show how closely the basic DTD expression matches d'Alembert's principle and how this closeness may help to solve the problems of complex chemical equilibria. The resemblance is in no way occasional: the theory of equilibrium, based on the balance of forces, inevitably hits d'Alembert's principle.

The major feature of our approach to complex chemical equilibrium is that Lagrangian (13) explicitly contains subsystem interactions as a factor, affecting equilibrium of the whole chemical system. These constraints in the lagrangian (13) express a demand that in global equilibrium all subsystems are supposed to be in equilibrium with their compliments to the mother system. Also, after all we don't need any more classical hypothesis of local equilibrium [9], which was often used as a crutch to imperfection of classical thermodynamics – now "local", or subsystem equilibria are naturally interwoven with global equilibrium. Global equilibrium in such a system is clearly a self-regulating state, based on the entangled and mutually guarding individual equilibria. No wonder, that being moved into "far-from equilibrium" area (beyond bifurcation point, [1,2]) quite regular at a glance objects show a plenty of non-classical behavior.

Another DTD feature is that we seek the equilibrium point in the space of subsystem deviations from their TdE, not in the space of chemical species, consistently treating the entity in question as a system, not just a set of chemical substances.

On historical and logical reasons, variational principles have been first formulated and achieved unsurpassed elegancy in analytical mechanics. Their successful usage even in a canonical, mechanical form in many other than mechanics areas means more than mere applicability - it proves time and again existence of common unique variational principles of nature, acting in all natural branches.

Results of this work can be used as a basis for algorithm to code software for thermodynamic simulation of complex chemical and similar systems. Expected advantage of potential DTD based software before the known classical simulation programs is at least a higher precision in finding the point of global equilibrium and in complex equilibrium compositions.